\documentstyle[12pt]{article}

\newcounter{eq}
\newcounter{sc}

\newcommand {\PL}   {Phys. Lett.}

\newcommand {\PR}   {Phys. Rev.}

\def\overleftrightarrow#1{\vbox{\ialign{##\crcr
 $\leftrightarrow$\crcr\noalign{\kern-1pt\nointerlineskip}
 $\hfil\displaystyle{#1}\hfil$\crcr}}}

\newlength{\minitwocolumn}
\begin{document}

\begin{flushright}
DPUR/TH/28\\
October, 2011\\
\end{flushright}
\vspace{20pt}

\pagestyle{empty}
\baselineskip15pt

\begin{center}
{\large\bf A Resolution to Cherenkov-like Radiation of OPERA Neutrinos
\vskip 1mm }

\vspace{20mm}
Ichiro Oda \footnote{E-mail address:\ ioda@phys.u-ryukyu.ac.jp}
and Hajime Taira

\vspace{5mm}
           Department of Physics, Faculty of Science, University of the 
           Ryukyus,\\
           Nishihara, Okinawa 903-0213, Japan.\\

\end{center}


\vspace{5mm}
\begin{abstract}
The OPERA collabotation has reported evidence of superluminal neutrinos
with a mean energy 17.5 GeV ranging up to 50 GeV.
However, the superluminal interpretation of the OPERA results has been recently 
refuted theoretically by Cherenkov-like radiation. 
We discuss a loophole of this argument from the kinematical viewpoint and
find it possible to avoid the Cherenkov-like radiation of the OPERA neutrinos.
The key idea of our argument is to admit the fact that the neutrinos travel faster than 
the observed speed of light while they do slower than the true speed of light in
vacuum so strictly speaking they are not superluminal but subluminal. Moreover, 
we present a model where these two velocities of light can be constructed by 
taking account of influences from dark matters near the earth.
\end{abstract}

\newpage
\pagestyle{plain}
\pagenumbering{arabic}


\rm
\section{Introduction}

The OPERA experiment has recently announced a remarkable discovery that
neutrinos from CERN to Gran Sasso travel faster than the speed of 
light \cite{OPERA}. Soon after this sensational announcement, there have been
a lot of activities to attempt to understand this interesting phenomenon of 
superluminal neutrions from the theoretical viewpoint. As a consequence of 
such theoretical studies, it has turned out that there are two theoretical 
challenges to the OPERA experimental results, which are the bremsstrahlung effects 
\cite{Cohen} and pion decay \cite{Gonzalez, Bi, Cowsik}. Since the two problems
stem from the same cause that the OPERA neutrinos are superluminal, we focus
on only the bremsstrahlung effects in this article. We will see that the
problem of pion decay is also automatically solved by our theory.

Let us start with an explanation of what we call, the bremsstrahlung effects or 
Cherenkov-like radiation of superluminal neutrinos. On the way from CERN to Gran Sasso, 
the very effect of superluminal propagation of neutrinos would have caused some 
distortions in the beam of neutrinos owing to Cherenkov-like radiation and severely 
depleted the higher-energy neutrinos, 
thereby making it impossible to observe neutrinos with more than 12.5 GeV energy. 
This theoretical result is obviously against the OPERA results where a lot of high-energy 
neutrinos above 12.5 GeV are observed as seen in the OPERA paper \cite{OPERA}
"Data were then split into two bins of nearly equal statistics, including events 
of energy lower or higher than 20 GeV. The mean energies of the two samples are 
13.9 and 42.9 GeV."

More recently, stimulated with the above theoretical observation, ICARUS group has
analyzed their data and found that "the neutrino energy distribution of the ICARUS
events in IAr agrees with the expectations from the Monte Calro predictions from
an unaffected energy distribution of beam from CERN. Our results therefore refute
a superluminal interpretation of the OPERA result according to the Cohen and
Glashow prediction for a weak current analog to Cherenkov radiation" \cite{ICARUS}.  

In order to make a physically plausible theory of the OPERA results, one has to
overcome this issue at all events. It is our aim in this article to present one solution
to this problem. To do so, it is worth keeping in mind that since physics of Cherenkov-like 
radiation of superluminal neutrinos is purely in the kinematical regime,
one has to propose a not dynamical but kinematical solution to this problem 
which would forbid this process from the kinematical reasons. 

In addition to it, we cannot help accepting the fact that according to Cohen and Glashow 
if a neutrino could travel faster than the speed of light in vacuum as claimed by OPERA, 
superluminal neutrinos rapidly lose energy via pair bremsstrahlung so few neutrinos with energies 
in excess of 12 GeV are detected at Gran Sasso.    

In this article, let us suppose that the OPERA results are correct at any rate
even if further experimental scrutiny is surely needed. Then, a question naturally arises:
"Is there a way out to reconcile the OPERA results which insist on the superluminal propagation
of neutrinos, with the bremsstrahlung effects which refute a superluminal interpretation of 
the OPERA results by the kinematics?"
Below we wish to present a solution to this question. The key idea is to introduce two
kinds of speeds of light in such a way that the OPERA neutrinos are not
$\it{superluminal}$ but $\it{subluminal}$ under the true velocity of light.

This article is organized as follows: In the next section, we will review the work 
by Cohen and Glashow. In Section 3, we explicitly make a model which has two speeds of 
light because of interaction between the gauge field and a Galileon-type of scalar field.
Section 4 is devoted to discussion.

\section{Review of bremsstrahlung effects and its resolution}

Let us begin with a review of the bremsstrahlung effects of superluminal
neutrinos \cite{Cohen}. This study is based on an old analysis of Ref. \cite{Coleman}
where many effects of violation of Lorentz invariance have been explored
in the framework of Standard Model.

The basic idea behind the work by Cohen and Glashow is that a neutrino traveling 
faster than the speed of light loses energy by emitting something owing to
weak interaction although neutrions do not carry electric charges (So this
process is dubbed Cherenkov-$\it{like}$-radiation as well). The most dominant decay
process of a muon neutrino, which mainly constitutes the OPERA beam, is found
to be $\nu_\mu \rightarrow \nu_\mu + e^+ + e^-$ whose threshold energy reads
\begin{eqnarray}
E =  \frac{2 m_e}{\sqrt{v_\nu^2 - 1}} \approx 140 MeV,
\label{Threshold}
\end{eqnarray}
for the OPERA neutrinos.

In the high energy limit where mass of an electron and neutrino can be
neglected, the rate of bremsstrahlung pair emission $\Gamma$ and its
associated rate of losing energy $\frac{d E}{d x}$ are calculated to be  
\begin{eqnarray}
\Gamma &=&  k' \frac{G_F^2}{192 \pi^3} E^5 ( v_\nu^2 - 1 )^3, \nonumber\\
\frac{d E}{d x} &=&  - k \frac{G_F^2}{192 \pi^3} E^6 ( v_\nu^2 - 1 )^3,
\label{Emission rate}
\end{eqnarray}
where $k = \frac{25}{448}, k' = \frac{1}{14}$ are numerical constants.
Then, the mean fractional loss of energy at each step of the bremsstrahlung
pair emission is given by
\begin{eqnarray}
\frac{- \frac{d E}{d x}}{\Gamma E} = \frac{k}{k'} \approx 0.78,
\label{Mean}
\end{eqnarray}
which implies that about $\frac{3}{4}$ of the neutrino energy is reduced
at one step of emission. 

Moreover, by integrating $\frac{d E}{d x}$ over a distance $L$, the final 
neutrino energy $E$ is expressed in terms of its initial one $E_0$ by
\begin{eqnarray}
\frac{1}{E^5} = \frac{1}{E_0^5} 
+ 5k \frac{G_F^2}{192 \pi^3} ( v_\nu^2 - 1 )^3 L.
\label{Final E}
\end{eqnarray}
By applying this formula to the case of the OPERA experiment, the
terminal energy of a neutrino at the OPERA detector is about 12.5 GeV.
In other words, few neutrinos reach the detector with energies in excess
of 12.5 GeV. Unfortunately, the OPERA detector observes neutrinos with 
the mean energy of 17.5 GeV ranging up to 50GeV \cite{OPERA}, so the calculation 
rules out an interpretation that the OPERA neutrinos are superluminal.
Incidentally, the ICARUS group has recently reanalyzed their data obtained
in 2010 and found that no Cherenkov-like radiation has been detected in 
ICARUS \cite{ICARUS}.

Where is there a loophole in the above argument? 
It is true that we cannot help accepting the conclusion by Cohen and
Glashow as long as a neutrino is superluminal. Thus, the only way out
is to consider that the OPERA neutrinos are not superluminal but
subluminal, but then how can we make them stay subluminal, which appears
to be against the OPERA report? 

As an implicit assumption of the argument by Cohen and Glashow, the speed of light 
is considered as a universal constant and does not receive any influences 
from the environment. 
However, it has been already pointed out that there is the possibility that 
the velocity of the photons might be dependent on energy from the HERA Compton 
polarization data \cite{Gharibyan}. \footnote{We identify the velocity of 
the photons with that of light.}
It might be therefore reasonable to conjecture that 
the speed of light would receive some influences from surroundings and consequently 
the observed speed of light in vacuum might become smaller than the true speed of 
light in vacuum.  

Put differently, if the group velocity $v_\nu$ of the OPERA neutrions is
between the observed velocity $c_0$ of light and the true one $c$, i.e., 
$c_0 < v_\nu < c$, the OPERA neutrions can be regarded as subluminal since the
property of superluminality or subluminality is defined by using the
true velocity $c$ of light in vacuum. 

Then, the problem of the bremsstrahlung effects is converted to a different problem: 
Can we construct a model which has two kinds of velocities of light,
which satisfy the relation $c_0 < c$, without conflicting
with special relativity? In the next section, we shall present such a
model based on the recent works \cite{Dvali, Iorio, Kehagias, Wang, Saridakis, Oda}.

\section{A model of two velocities of light}

We shall begin with the Lagrangian density of our theory : \footnote{We make use of 
the flat metric $\eta_{\mu\nu} = diag ( -1, +1, +1, +1)$ for raising or lowering indices.}
\begin{eqnarray}
{\cal{L}} = - \frac{1}{4} F_{\mu\nu} F^{\mu\nu} 
+ \frac{1}{2 M_*^4} \partial^\nu \pi \partial^\alpha \pi F_{\mu\nu} F^\mu \ _\alpha,
\label{Original Action}
\end{eqnarray}
where $M_*$ is a mass scale which controls the strength of the coupling between
the Galileon-type of scalar field $\pi$ \cite{Nicolis} and the abelian gauge field $A_\mu$.
The gauge field strength $F_{\mu\nu}$ is defined by $F_{\mu\nu} = \partial_\mu A_\nu
- \partial_\nu A_\mu$ as usual.  In the absence of the $\pi$ field, the photon
travels at the speed of light in vacuum since the gauge field satisfies the conventional
Maxwell equations. In the theory at hand, it is essential to define
this velocity, which we denotes as $c$, as the $\it{true}$ speed of light in vacuum, 
which appears in various formulae in special relativity. Thus, the causal structure 
should be determined on the basis of this velocity $c$ and the flat metric $\eta_{\mu\nu}$.

Now it is easy to rewrite (\ref{Original Action}) as
\begin{eqnarray}
{\cal{L}} = - \frac{1}{4} ( \eta^{\mu\alpha} - \frac{1}{M_*^4} \partial^\mu \pi 
\partial^\alpha \pi ) ( \eta^{\nu\beta} - \frac{1}{M_*^4} \partial^\nu \pi 
\partial^\beta \pi ) F_{\mu\nu} F_{\alpha\beta},
\label{Original Action 2}
\end{eqnarray}
thereby making it possible to read out an effective metric
\begin{eqnarray}
g^{\mu\nu}_{(A)} = \eta^{\mu\nu} - \frac{1}{M_*^4} \partial^\mu \pi
\partial^\nu \pi, 
\label{E-metric}
\end{eqnarray}
along which the photon propagates. 
 
Next, let us consider the spherical symmetric configuration for the scalar
\begin{eqnarray}
\pi = \frac{\alpha}{r}, 
\label{Scalar}
\end{eqnarray}
where $\alpha$ is a constant \cite{Kehagias}. Note that this configuration
breaks the Lorentz invariance spontaneously. Then, it turns out that 
the effective space-time on which the photon propagates has the line element
\begin{eqnarray}
ds^2 \equiv g_{(A) \mu\nu} d x^\mu d x^\nu 
= - dt^2 + \frac{1}{1 - \frac{\alpha^2}{M_*^4} \frac{1}{r^4}} dr^2
+ r^2 d \Omega^2_2,
\label{Line}
\end{eqnarray}
where $d \Omega^2_2 \equiv d \theta^2 + \sin^2 \theta d \phi^2$.

It is then straightforward to derive an effective velocity $c(r)$ of the photon
for the fixed angles $(\theta, \phi)$
\begin{eqnarray}
c(r) \equiv \frac{dr}{dt} = \left( 1 - \frac{\alpha^2}{M_*^4} 
\frac{1}{r^4} \right)^{\frac{1}{2}} c,
\label{E-velocity}
\end{eqnarray}
where for conveniece the $\it{true}$ velocity $c$ of light in vacuum is recovered.
Note that at the spatial infinity $r \rightarrow \infty$, the effective velocity
coincides with the $\it{true}$ velocity $c$, that is, 
\begin{eqnarray}
\lim_{r \to \infty} c(r) = c.
\label{E-velocity 2}
\end{eqnarray}
Furthermore, note that the difference between $c$ and $c(r)$ is positive-definite
as long as the constant $\alpha$ is non-zero
\begin{eqnarray}
c - c(r) = \frac{\alpha^2}{2 M_*^4} \frac{1}{r^4} > 0,
\label{E-velocity 3}
\end{eqnarray}
which holds when $\frac{\alpha^2}{M_*^4} \frac{1}{r^4} \ll 1$.

On our earth, the observed velocity of light is given by $c(r)$ when the photons
are located at the place whose distance in the radial direction is $r$ from the center 
of the earth. If we assume that the speed of a neutrino, denoted as $v_\nu$, takes
a definite value and is smaller than $c$ but larger than $c(r)$ 
\begin{eqnarray}
c(r) < v_\nu < c,
\label{Velocity ansatz}
\end{eqnarray}
we would have a superluminal neutrino for the velocity $c(r)$ as observed in the
OPERA experiment
\begin{eqnarray}
\beta(r) \equiv \frac{v_\nu - c(r)}{c(r)} > 0,
\label{Delta(r)}
\end{eqnarray}
whereas we have a subluminal neutrino for the true velocity $c$
\begin{eqnarray}
\beta \equiv \frac{v_\nu - c}{c} < 0.
\label{Delta}
\end{eqnarray}

Recalling that the property of superluminality or subluminality of neutrinos
is now defined by using the $\it{true}$ velocity $c$ of light in vacuum,
the OPERA neutrinos are actually not superluminal but subluminal! Hence,
we do not have Cherenkov-like radiation for the OPERA neutrinos at all
since we can always take the rest frame for the subluminal neutrino. 
This is our resolution to the problem of the bremsstrahlung effects of 
the OPERA neutrinos.
It is worthwhile to notice that our solution is purely kinematical as 
desired. \footnote{Recently, a similar resolution has been also proposed 
by Nakanishi in Ref. \cite{Nakanishi}. It is a pity that his interesting
article is written only in Japanese.}

Let us show that the results of OPERA and SN1987A give us information
on the mass scale $M_*$. First, let us note that the OPERA result yields
a condition for the dimensionless quantity $\beta(r)$
\begin{eqnarray}
\beta(R_\oplus) \equiv  \frac{v_\nu - c(R_\oplus)}{c(R_\oplus)} \approx 2.5 \times 10^{-5},
\label{Beta}
\end{eqnarray}
where $R_\oplus$ is the radius of the earth and takes the value $R_\oplus = 6.4 
\times 10^8 cm$.

Next, let us utilize the fact that neutrinos from SN1987A to the earth travels
at almost the same velocity as the true velocity of light in vacuum, so we have a relation
\begin{eqnarray}
v_\nu \approx c.
\label{SN1987A}
\end{eqnarray}
With the help of this relation (\ref{SN1987A}), Eq. (\ref{Beta}) can be cast to the form
\begin{eqnarray}
\beta(R_\oplus) \approx  \frac{\alpha^2}{2 M_*^4} \frac{1}{R_\oplus^4}.
\label{Beta 2}
\end{eqnarray}
At this stage, we may assume the scalar field $\pi$ is sourced by the trace
part of the energy-momentum tensor on the earth which does not include the contribution
from the scalar field $\pi$ itself. With this assumption, the numerical constant
takes the form in the static case \cite{Kehagias}
\begin{eqnarray}
\alpha = \frac{M_\oplus}{M_*} \equiv M_\oplus L_*,
\label{Alpha}
\end{eqnarray}
where $M_\oplus$ is the mass of the earth and we have defined the length scale
by $L_* = \frac{1}{M_*}$.

Using Eq's. (\ref{Beta}),  (\ref{Beta 2}) and (\ref{Alpha}),
$L_*$ is calculated as
\begin{eqnarray}
L_* \approx  \left( 2 \times 10^{-4} L_{Pl}^4 \frac{R_\oplus^4}{R_{\oplus SS}^2}
\right)^{\frac{1}{6}} \approx 3 \times 10^{-17} cm,
\label{L_*}
\end{eqnarray}
where $L_{Pl}, R_{\oplus SS}$ are respectively the Planck length and the Schwarzschild
radius of the earth, and are explicitly given by $L_{Pl} = 1.6 \times 10^{-33} cm,
R_{\oplus SS} = 0.89 cm$. Then, it is of interest to notice that $M_*$ is approximately 
equivalent to the energy scale where new physics beyond Standard Model appears
\begin{eqnarray}
M_* \approx  1 TeV,
\label{TeV}
\end{eqnarray}
above which the coupling between the gauge field and the scalar field becomes strong.

In this context, it is tempting to identify the $\pi$ scalar field with a
candidate of dark matters. Actually, this identification seems to be
consistent with the results of OPERA and SN1987A at the same time by
the following reasoning: In general, dark matters are expected to have 
a tendency of localizing near massive objects such as stars and the earth 
owing to gravitational interaction, compared to empty regions of outer space.
The reason why neutrinos from SN1987A to the earth traveled at almost the same 
velocity as the true velocity of light in vacuum is that since there are not 
so much  dark matters in outer space, the neutrinos from SN1987A propagated 
at the true velocity $c$ without interacting with dark matters. 
On the other hand, since it is expected that there are sufficient dark matters 
on the earth, the interaction between the photons and dark matters reduces 
the speed of light on the earth to the smaller observed velocity $c_0$ from
the larger true velocity $c$.

\section{Discussion}

In this article, we have presented a resolution to one serious theoretical problem,
that is, the bremsstrahlung effects of superluminal neutrions. Our resolution
is on the kinematical grounds and thus strictly forbids the OPERA neutrinos to
emit a pair of electron and positron via the bremsstrahlung effects.
In this article, we have introduced the Galileon-type of scalar field to interact
with the gauge field, but it would be possible to consider the other 
fields such as spin 2 new tensor field and spin 1/2 spinor field instead of the
scalar field.

It is then natural to ask ourselves if our resolution also provides a resolution to the 
other problems associated with the OPERA results. In particular, it is now known that
the other challenging theoretical issue lies in pion decay process  \cite{Gonzalez, Bi, Cowsik}
where it was found that the decay of charged pion $\pi^+ \rightarrow \mu^+ + \nu_\mu$,
which is nothing but the neutrino production process in the OPERA experiment,
becomes kinematically forbidden for $E_\nu > 5 GeV$, which is obviously inconsistent
with the OPERA results. It is worth stressing that our resolution also provides a solution 
to this problem since this decay process is automatically prohibitted whenever the OPERA 
neutrinos are subluminal.

Finally, let us comment on the connection with related works. Moffat has presented
a similar idea by taking account of bimetric relativity \cite{Moffat}. 
Li and Nanopoulos have recently advocated an idea that all the Standard
Model particles might be subluminal due to background effects on the basis of
a string theory-inspired model \cite{Li}.  Even if our present theory is 
very different from these works, it might be interesting to investigate
the relation more closely in future.

\begin{flushleft}
{\bf Acknowledgements}
\end{flushleft}

This work (I.O.) is supported in part by the Grant-in-Aid for Scientific 
Research (C) No. 22540287 from the Japan Ministry of Education, Culture, 
Sports, Science and Technology.


\end{document}